\pdfoutput=1
\documentclass[aps,prb,a4paper,10pt,twocolumn,superscriptaddress]{revtex4-1}

\usepackage{graphicx}
\usepackage[multi-part-units = single]{siunitx}
\usepackage{amsmath}

\begin{document}

\title{Tracing the phase of focused broadband laser pulses}

\author{Dominik Hoff}
\altaffiliation{These authors contributed equally to this work.}
\affiliation{Helmholtz-Institut Jena and Institut f\"ur Optik und Quantenelektronik, 
Friedrich-Schiller-Universit\"at Jena, 
\small Max-Wien-Platz~1, D-07743 Jena, Germany}

\author{Michael Kr{\"u}ger}
\altaffiliation{These authors contributed equally to this work.}
\affiliation{Department Physik, Friedrich-Alexander-Universit\"at Erlangen-N\"urnberg (FAU),
\small Staudtstr.~1, D-91058 Erlangen, Germany}
\affiliation{Department of Physics of Complex Systems, Weizmann Institute of Science,
\small 234 Herzl St., Rehovot 76100, Israel}

\author{Lothar Maisenbacher}
\affiliation{Max-Planck-Institut f\"ur Quantenoptik, \small Hans-Kopfermann-Str.~1, D-85748
Garching, Germany}

\author{A. M. Sayler}
\affiliation{Helmholtz-Institut Jena and
Institut f\"ur Optik und Quantenelektronik, Friedrich-Schiller-Universit\"at Jena, \small 
Max-Wien-Platz~1, D-07743 Jena, Germany}

\author{Gerhard G. Paulus}
\affiliation{Helmholtz-Institut Jena and
Institut f\"ur Optik und Quantenelektronik, Friedrich-Schiller-Universit\"at Jena, \small 
Max-Wien-Platz~1, D-07743 Jena, Germany}

\author{Peter Hommelhoff}
\affiliation{Department Physik, Friedrich-Alexander-Universit\"at Erlangen-N\"urnberg (FAU),
\small Staudtstr.~1, D-91058 Erlangen, Germany}
\affiliation{Max-Planck-Institut f\"ur Quantenoptik, \small Hans-Kopfermann-Str.~1, D-85748
Garching, Germany}
\affiliation{Max-Planck-Institut f\"ur die Physik des Lichts, \small Staudtstr.~2,
D-91058 Erlangen, Germany}

% \title{The original paper is available at http://dx.doi.org/10.1038/nphys4185.}
% \footnote{The original paper is available at \url{http://dx.doi.org/10.1038/nphys4185}.}
% \date{\today}

\maketitle

The corresponding Nature Physics paper and the supplementary material is available at 
\url{http://dx.doi.org/10.1038/nphys4185} .

%%%%%%%%%%%%%%%%%%%%%%%%%%%%%%%%%%%%%%%%%%%%%%%%%%%%%%%%%%%%%%%%%%%%%%%%%%%%%%%%%%%%%%%%%%

{\bf
Precise knowledge of the behaviour of the phase of light in a focused beam is fundamental to
understanding and controlling laser-driven processes.
More than a hundred years ago an axial phase anomaly for focused monochromatic light beams was 
discovered and is now commonly known as the Gouy phase \cite{Gouy1890, Born1999, Siegman1986, 
Visser2010}.
Recent theoretical work has brought into question the validity of applying this 
\textit{monochromatic} phase formulation to the broadband pulses becoming ubiquitous today 
\cite{Porras2002, Porras2009}.
Based on electron back-scattering at sharp nanometre-scale metal tips, a method is available to 
measure light fields with sub-wavelength spatial resolution and sub-optical cycle time resolution 
\cite{Kruger2011, Wachter2012, Piglosiewicz2014}.
Here we report such a direct, three-dimensional measurement of the spatial dependence of the
optical phase of a focused, $4$-fs, near-infrared laser pulse.
The observed optical phase deviates substantially from the monochromatic Gouy phase ---
exhibiting a much more complex spatial dependence, both along the propagation axis and in the radial
direction.
In our measurements, these significant deviations are the rule and not the exception for focused, 
broadband laser pulses.
Therefore, we expect wide ramifications for all broadband laser-matter interactions, such as in 
high-harmonic and attosecond pulse generation, femtochemistry \cite{Gordon2007}, 
ophthalmological optical coherence tomography \cite{Johnson2001, Drexler2008} and light-wave 
electronics \cite{Krausz2014}.
}

%%%%%%%%%%%%%%%%%%%%%%%%%%%%%%%%%%%%%%%%%%%%%%%%%%%%%%%%%%%%%%%%%%%%%%%%%%%%%%%%%%%%%%%%%%

In ultrafast light-matter interactions, the phase of the optical carrier field with respect to the
pulse envelope's maximum --- the carrier-envelope phase (CEP, see Methods) \cite{Udem2002} --- is
one of the fundamental controls that allows one to steer chemical reactions \cite{Kling2006}, the
generation of attosecond pulses via high-harmonic generation \cite{Baltuska2003}, and electron
emission and acceleration from solid surfaces and nanostructures \cite{Zherebtsov2011, Kruger2011,
Piglosiewicz2014}, among others.
Hence, determining and controlling the CEP is mandatory in many fields using lasers, but taking
into account the broadband and often intense and ultrashort nature of these pulses is challenging 
and an active area of research \cite{Fordell2011, PaaschColberg2014}.
Further, nonlinear light-matter interactions usually take place in the focus of a beam where the
CEP shows a strong spatial dependence. 
Thus, for a detailed understanding of and field control over these processes, it is essential to 
take the focal phase evolution, target position, and target extent into account
\cite{Goulielmakis2007, Krausz2014}.
This is as important as controlling the CEP of the input beam itself.

For a focused monochromatic beam, the on-axis phase shift due to diffraction is described by the 
familiar Gouy phase, which follows a simple arctangent curve \cite{Gouy1890, Born1999}.
However, many of today's coherent light sources, even some as common place as those used in 
ophthalmological diagnostics, are far from being monochromatic.
Rather, they can span close to, and many even exceed, an octave of spectral bandwidth 
\cite{Drexler2008, Shverdin2005, Wirth2011}.
Further, recent theoretical studies, based on diffraction theory for pulsed Gaussian beams,
yielded spatially-dependent phases that significantly deviate from the simple Gouy phase and show a
much more complex behaviour that is dictated by the wavelength-dependent input beam geometry
\cite{Porras2002, Porras2009}.
The need for further investigation is underscored by experimental studies, which strongly suggest 
deviations from the monochromatic Gouy phase \cite{Lindner2004, Tritschler2005, Major2015a}.

\begin{figure*}[htb]
\includegraphics[width=12cm]{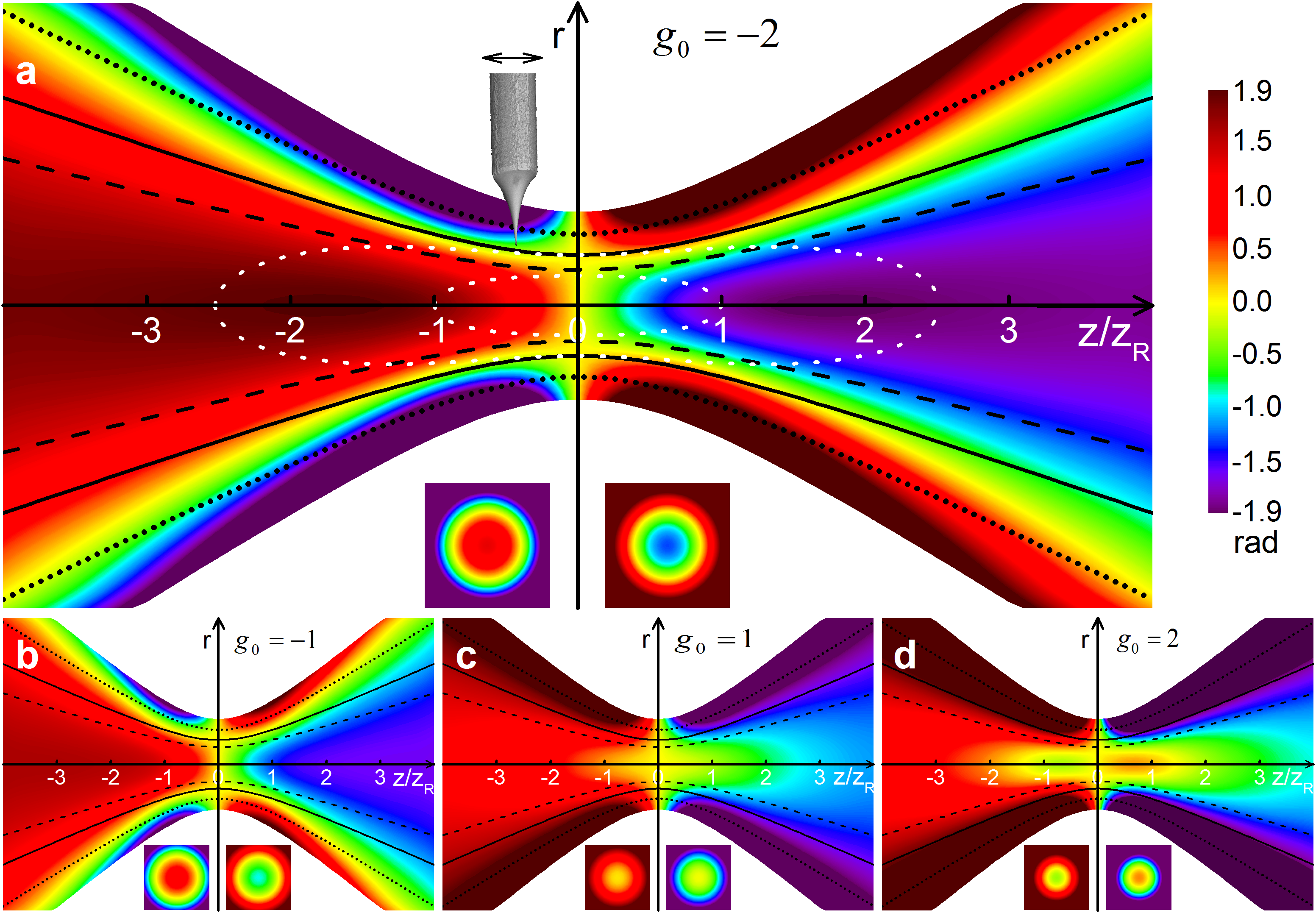}
\caption{
{\bf Carrier-envelope phase in a focused, broadband, pulsed Gaussian beam.}
The colour plots show the calculated CEP in the focus of an input beam with geometry
factors of ({\bf a}) $g_0 = -2$, ({\bf b}) $g_0 = -1$, i.e. isodivergent, ({\bf c}) $g_0 = 1$,
which has a wavelength independent beam waist, and ({\bf d}) $g_0 = 2$ as a function of axial and
radial position, $z$ and $r$ respectively.
The factor $g_0$ is explained in the text and describes the spectral input beam geometry.
For visual convenience, the colour is truncated for $|\Delta\phi| > \SI{1.9}{rad}$ and the beam 
size 
at $r > 1.86~\mathrm{w}(z)$ where the intensity has dropped below a factor of $10^{-3}$ of the 
on-axis value.
Cross sections in the $x$-$y$-plane are given for an alternative perspective of the same data at 
$z= -0.5$ and $0.5~ z_R$ in the same radius range (insets).
The black lines are hyperbolae related to the local $1/e^2$ intensity radius $\mathrm {w}(z)$, 
namely at  $r = \mathrm {w}(z) / \sqrt {2}$ (dashed), $r = \mathrm{w}(z)$ (solid), $r = \sqrt {2}~ 
\mathrm{w}(z)$ (dotted).
The white dotted lines are isointensity curves of $I_0/2$ and $I_0/e^2$ (see Supplementary
information).
The CEP is probed by recording photoelectron spectra with a metal nanotip (drawn to scale) on the 
optical axis and on the hyperbolae in the half space towards the tip.
In the measurements the Rayleigh length and waist radius are approximately $\SI{400(50)}{\um}$ and 
$\SI{9 \pm 0.5}{\um}$, respectively.
}
\label{fig:2DFScheme2}
\end{figure*}
In recent years, significant advancements have been made in CEP detection and control
\cite{Baltuska2003, Udem2002, Wittmann2009, Fordell2011}, which have facilitated and driven the
discovery of more and more processes that are dependent on and can be controlled via the CEP.
One such phenomenon is the strong-field, few-cycle-laser-driven back-scattering of photoelectrons, 
i.e. those electrons freed by the laser field that can then be driven back to and scattered off 
their parent matter when the field flips sign within an optical cycle.
The large kinetic energy obtained by these electrons strongly depends on the CEP, which we utilize 
as an experimental signature, see Supplementary information.
This extremely CEP-sensitive effect was observed in noble gases and recently also at solid state 
nanotips \cite{Kruger2011, Wachter2012, Piglosiewicz2014}.
For the latter, strong-field induced photo-emission happens almost exclusively in the enhanced
optical near-field region at the apex of the sharp tip, with a radius of $\sim \SI{10}{nm}$
\cite{Thomas2013}.
Thus, electrons from such a highly-localized source are particularly well suited to be used as a
sensor to probe focused light fields with resolution better than their natural length and time 
scales.
Namely here, with a spatial resolution of $\sim \SI{10}{nm}$ ($\ll \lambda/2 \approx 350\,$nm, the 
typical length scale of a focus of light with wavelength of $\lambda \approx 700$\,nm), and a 
CEP resolution of $\sim \SI{80}{mrad}$, corresponding to $\sim 60$ attoseconds ($\ll 2.3\,$fs, the 
optical period of light with $\lambda \approx 700\,$nm).

\begin{figure}[htb]
\begin{center}
\includegraphics[width=8.5cm]{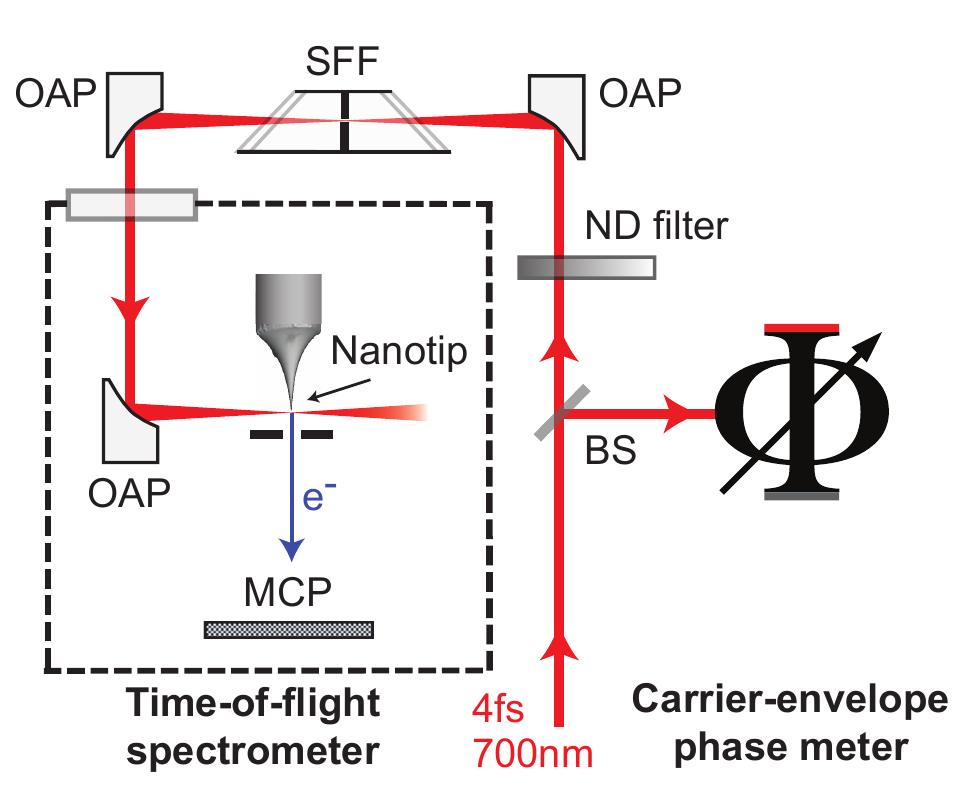}
\end{center}
\caption{
{\bf Overview of the phase-tagging scheme with a nanotip.}
For parallel determination of the laser pulse's CEP and the time-of-flight of electrons from the 
nanotip, an amplified $\SI{4}{fs}$, $\SI{700}{nm}$ central wavelength, horizontally polarized pulse 
train from a laser system with a hollow core fibre compressor is split at the beam splitter 
{\bf(BS)} towards the carrier-envelope phase meter arm (right), and the time-of-flight spectrometer 
arm (left), respectively.
The spatial mode of the beam to the nanotip is cleaned with a spatial frequency filter {\bf(SFF)}.
The beam is focused inside the ultrahigh vacuum chamber with a \SI{90}{\degree} off-axis parabolic
mirror ({\bf OAP}, $f = \SI{50}{mm}$) on the nanotip.
The TOF spectrometer records the flight time of photoelectrons with a microchannel-plate {\bf(MCP)}
that is then converted into kinetic energy and provides the x-axis for the spectrum in 
Fig. S2, see Supplementary information.
The CEP-meter measures the randomly varying carrier-envelope phase of each and every shot and thus 
resolves the spectrum along the y-axis in Fig. S2.
For full details see Supplementary information.
}
\label{fig:setup}
\end{figure}

In this work we present a quantitative, direct, three-dimensional mapping of the CEP evolution of a
focused broadband laser beam, spanning a range of seven times the Rayleigh range along the
propagation axis and one and a half times the local beam radius perpendicular to the optical axis.
To achieve this, we have combined a nanotip-based focus characterisation setup with a 
xenon-gas-based, stereographic, above-threshold ionisation (ATI) CEP-meter.
Every laser pulse from a hollow-core fibre compressor is split between the two separate, but 
synchronized and parallel measurements: one for individual characterisation of the 
random CEP of each shot and the other for recording electron spectra from the 
laser-nanotip interaction \cite{Wittmann2009, Johnson2011}, see Methods and 
Supplementary information.
In this way, every detected electron from the nanotip can be associated with the 
CEP-value for the specific laser pulse producing that electron.
The beam towards the nanotip focus measurement was spatially filtered to remove higher 
modes and provide a good TEM$_{00}$ approximation as required for the following model.
Measurements were done to probe the relative CEP-shift both along the laser's 
propagation axis and radially outward, mapping the cylindrically-symmetric three-dimensional space.
Both tungsten and gold tips were used.
In comparison to previous experimental work \cite{Lindner2004, Tritschler2005, Major2015a}, in this 
measurement we probe the focus of a typical intense few-cycle pulse laser beam with a significantly 
better spatial resolution, scan a larger range of the focal volume and attain a well-defined focus 
with minimized aberrations by use of an off-axis parabolic mirror and a spatial filter element 
upfront.
In combination with considerable efforts to flatten the spectral phase, these factors allow for a 
clear and thorough experimental and theoretical investigation of deviations from the monochromatic 
Gouy phase and additionally reveal the transverse phase structure.

The longitudinal phase difference for a focused, \textit{continuous} and \textit{monochromatic}
Gaussian beam as compared to a plane wave is described along the propagation axis by

\begin{align}
 \Delta\psi (z) = -\arctan\left(\frac{z}{z_\mathrm{R}}\right),
\label{eq:gouy}
\end{align}
and is known as the Gouy phase, where $z$ is the laser propagation direction and
$z_\mathrm{R}$ is the Rayleigh length \cite{Gouy1890, Lindner2004, Porras2009}.
Laser \textit{pulses} on the other hand are composed of
a coherent superposition of a broad range of wavelengths, leading to a much more complex, 
spatially-dependent phase profile.
Moreover, the transverse shape and the divergence of the Gaussian beam in front of the focusing 
element are, in general, wavelength dependent.
Thus, it is necessary to take into account the wavelength-dependent input beam geometry in order to
determine the spatial dependence of the CEP after the focusing element.
A more generalized treatment of the strong focusing of chirp-free pulsed Gaussian 
beams in terms of their properties prior to focusing, combines the concept of enveloped carrier 
oscillations with fundamental diffraction theory \cite{Born1999}, which introduces iso-carrier-phase 
fronts and pulse-peak fronts \cite{Porras2002, Porras2009}. 
Their difference results in a relative CEP-shift in the focal area of

\begin{align}
\Delta\phi (z, r) =
-\arctan\left(\frac{z}{z_\mathrm{R}}\right)+ \frac{g_{0} \cdot 
\left[1-2\left(\frac{r}{\mathrm{w}(z)}\right)^{2} \right]}{\frac{z}{z_\mathrm{R}} 
+\frac{z_\mathrm{R}}{z}} ,
\label{eq:dphi}
\end{align}
where $\Delta\phi$ is defined to be $0$ in the focal reference plane (defined by $z=0$) for all 
radii $r$; here, $z_\mathrm{R}$ is the Rayleigh length at the centre frequency;
$\mathrm{w}(z)$ is the $z$-dependent beam radius, and $r/\mathrm{w}(z)$ is the normalized radial
coordinate, see Supplementary information.
The last term of Eq.~\ref{eq:dphi} describes the difference to the axial Gouy phase, 
Eq.~\ref{eq:gouy}, and accounts for the wavelength-dependent geometry of the input beam and 
the difference in curvature of the carrier-phase fronts as compared to that of the pulse fronts.
This term scales with $g_0$ --- a dimensionless geometry factor of the input beam evaluated at the
central angular frequency of the laser spectrum $\omega_0$, which we call the Porras factor
\cite{Porras2002, Porras2009}.

\begin{figure}
\includegraphics[width=8.5cm]{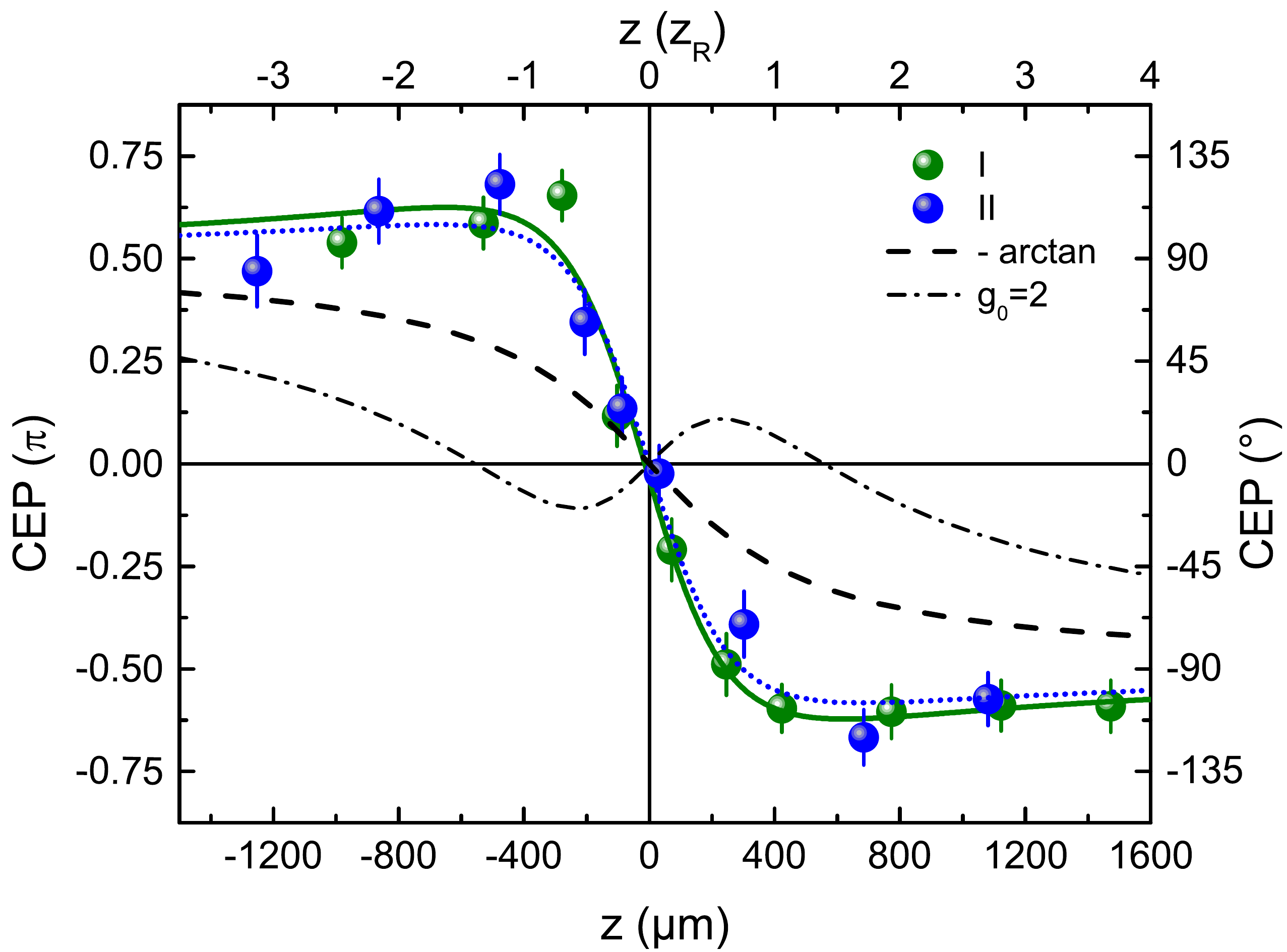}
\caption{
{\bf On-axis carrier-envelope phase.} CEP as a function of propagation distance, $z$, on
the optical axis.
Two traces have been recorded, data set I (green spheres, Rayleigh length $z_\mathrm{R}=380 \pm 
50~\si{\um}$) and II (blue spheres, $z_\mathrm{R}=365 \pm 50~\si{\um}$). 
The error bars show the measurement uncertainties in CEP of $\approx \SI{0.2}{\radian}$, as 
described in the Supplementary information. 
The experimental data points were fitted by a theoretical model (Eq.~\ref{eq:dphi} solid green and 
dotted blue curves), yielding fitting parameters of $g_0 = -2.1 \pm 0.2$ (reduced $\chi^2=1.3$) 
for data set I and $g_0 = -1.8 \pm 0.3$ (reduced $\chi^2=1.6$) for data set II. 
For comparison, we show the monochromatic Gouy phase, corresponding to the case $g_0 = 0$, for a
Rayleigh length of $z_\mathrm{R} = 400~\si{\um}$ (black dashed curve), which is clearly unable to 
explain our observations.
The dashed-dotted line depicts the theoretical curve for $g_0 = +2$ and $z_\mathrm{R} = 
400~\si{\um}$, as an example phase behaviour for another beam geometry (lineout of 
Fig.~\ref{fig:2DFScheme2}~d).
Data sets I and II have been recorded with different tip materials, see Supplementary information.
}
\label{fig:WAu}
\end{figure}

The Porras factor is given by
\begin{align}
g_0 = \frac{dZ_{\mathrm{R}}(\omega)}{d\omega} \bigg|_{\omega_0}\cdot
\frac{\omega_0}{Z_{\mathrm{R}}(\omega_0)}~~
\label{eq:g0}
\end{align}
and represents the normalized first derivative of the \textit{input} beam's 
Rayleigh length, $Z_{\mathrm{R}}(\omega)$ (capital letters are used for input 
beam parameters), with respect to the laser's spectral angular frequencies, $\omega$, evaluated at 
$\omega_0$.
The Rayleigh length before the focussing element is linked to the frequency dependent input beam
waist, $\mathrm{W(\omega)}$, by $Z_\mathrm{R}(\omega) = \omega \cdot \mathrm{W(\omega)}^2/(2c)$, 
where $c$ is the speed of light.
Three characteristic cases can be highlighted:
(i) $g_0=-1$ for an isodiverging input beam, i.e. with a constant divergence angle for all $\omega$;
(ii) $g_0=0$ for an iso\-diffracting beam, i.e. with a constant Rayleigh length;
and (iii) $g_0=+1$ for a beam with a frequency independent waist radius \cite{Porras2009} (see Fig.
\ref{fig:2DFScheme2} for example phase profiles and Fig. \ref{fig:WAu} for case $g_0=0$).
Note that at $g_0=0$, all frequencies are diffracted identically in the first order approximation 
and hence the phase shift consists merely of the Gouy component.
However, one ought not simply assume or expect $g_0$ to be $0$ as this is just one special and 
specific case, which is within a continuous range of possible $g_0$ values and rarely found in 
practice.

%%%%%%%%%%%%%%%%%%%%%%%%%%%%%%%%%%%%%%%%%%%%%%%%%%%%%%%%%%%%%%%%%%%%%%%%%%%%%%%%%%%%%%%%%%

Figure \ref{fig:WAu} shows the results of two on-axis measurements together with the monochromatic 
Gouy phase curve.
The measured CEP shows extrema at $z \approx 1.7$ times the Rayleigh length both before and
after the focus, resulting in a much steeper slope in the focal plane as compared to the
arctangent-curve of the Gouy phase.
The data can be fitted well by Eq.~\ref{eq:dphi} at $r=0$, resulting in $g_0=-2.1 \pm 0.2$ 
for the first (I) and $-1.8 \pm 0.3$ for the second (II) dataset.
These $g_0$-values are corroborated by independent measurements of the spectral input beam 
properties outside the vacuum chamber, namely the spectrally resolved Rayleigh 
length calculated from the beam diameter. This provides an attractive and more easily attainable 
alternative way to estimate $g_0$ via Eq.~\ref{eq:g0} and, therefore, obtain an approximate idea of 
the focal CEP evolution in the interaction region, see Supplementary information and 
Fig.~\ref{fig:2DFScheme2}.
Note that further work is warranted and needs to determine whether this technique 
will fulfill its promise.
In addition, the similarity of the experimental curves in Fig. \ref{fig:WAu} shows the
reproducibility of the measurement. 

\begin{figure}[htb]
\includegraphics[width=8.5cm,keepaspectratio=true]{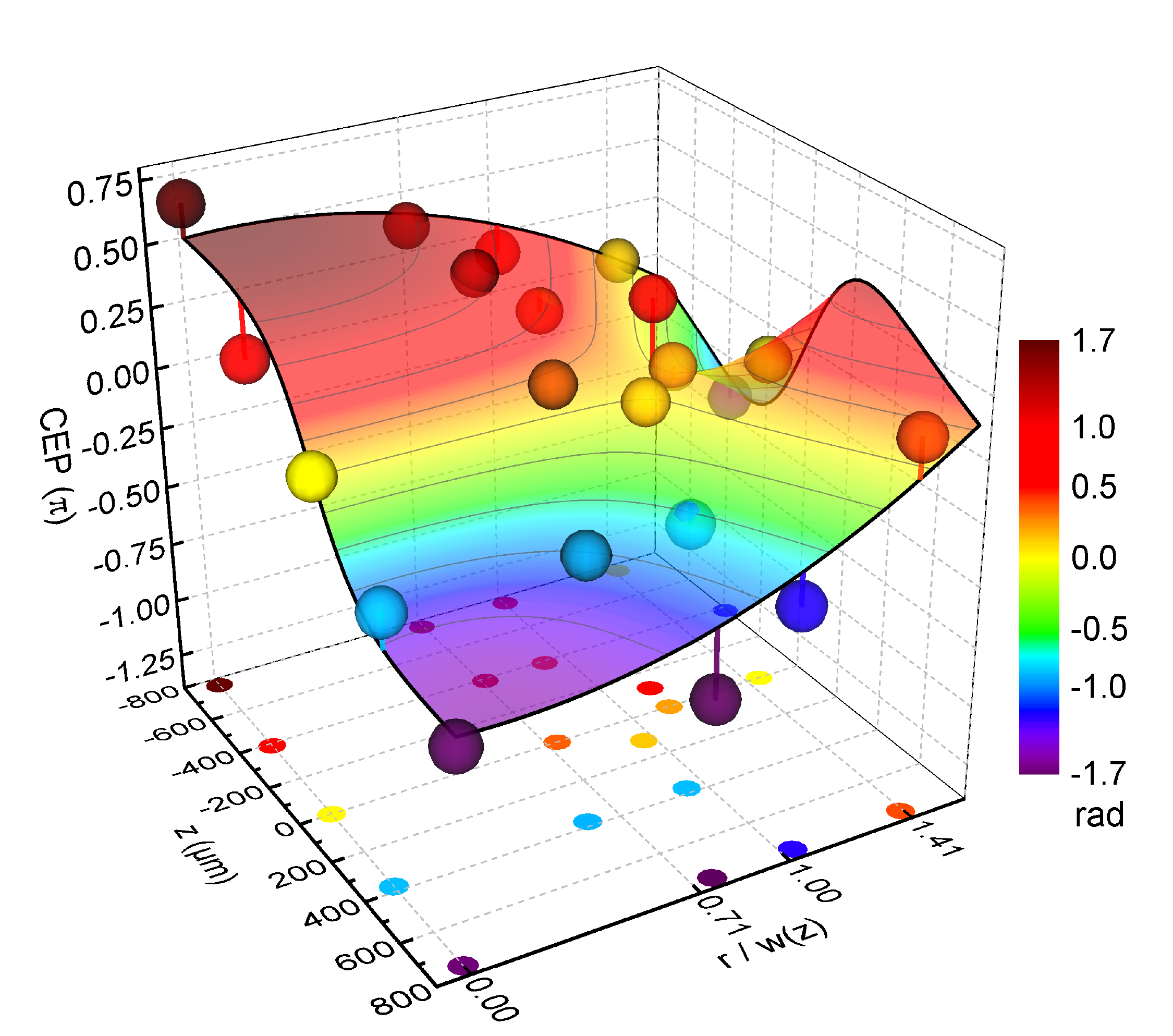}
\caption{
{\bf Off-axis carrier-envelope phase evolution.}
Three-dimensional plot of the CEP as a function of propagation distance $z$ and normalized radial
coordinate $r / \mathrm{w}(z)$.
The experimental data points are depicted by colour-coded spheres whose size encompasses their 
uncertainty resulting from measurement errors in positioning and determination of the CEP ($\approx 
\SI{0.2}{\radian}$), as described in the Supplementary information.
To further clarify the $z$-$r$-position of the measured data, each data point is projected in the 
$z$-$r$-plane.
A surface defined by Eq.~(\ref{eq:dphi}) was fit to the data (coloured), yielding a fit 
parameter of $g_0 = -1.2 \pm 0.3$ (reduced $\chi^2=3.7$). 
See also Fig. \ref{fig:2DFScheme2}b for a plot with a similar $g_0$ and 
Supplementary information for a stereoscopic view and a projected contour plot.
The Rayleigh length is $\SI{350(50)}{\um}$, independently determined with a knife-edge measurement.
}
\label{fig:2DPunkte}
\end{figure}

We were also able to produce a three-dimensional map of the CEP by scanning both $z$ and $r$ with
the tip (Fig. \ref{fig:2DPunkte}).
We observe that there is a strong radial dependence of the CEP, again differing from the
monochromatic Gouy phase.
Fitting Eq.~\ref{eq:dphi} to the data points yields $g_0 = -1.2 \pm 0.3$, where $g_0$ and 
the offsets in $z, r$ and $\Delta\phi$ are free parameters and the Rayleigh length $z_\mathrm{R}$ 
was determined independently by a knife-edge measurement to $z_\mathrm{R}=\SI{350(50)}{\um}$, see 
Supplementary information.
For negative $g_0$ values, as observed in all measurements here, characteristic features of the CEP
around the focus can be recognized, see Figs. \ref{fig:2DPunkte} and \ref{fig:2DFScheme2}:
(i) The radial dependence of the CEP surface is concave before, e.g. at $z=\SI{-800}{\um}$, and
convex after the focus, e.g. at $z=\SI{800}{\um}$.
(ii) Radially further out from the axis, i.e. $r\gtrsim \mathrm{w}(z)$, the CEP changes sign, 
implying that the vector potential at the pulse peak changes sign as well (see Supplementary Eq.~9).
In other words, the CEP evolves with opposite slopes on and off the optical axis.
And (iii) in the focal plane at the waist radius, i.e.\ at $(z, r) = (0, \mathrm{w_0})$, the
phase exhibits a saddle point.
In this measurement, the value for $g_0$ differs from that obtained previously in the on-axis case 
as it was done with different laser tuning.
This underscores the need to properly characterise the light source in use and determine the 
influence of laser tuning parameters, such as hollow-core fibre pressure, on the spectral geometry 
of the output beam and $g_0$ in order to facilitate advanced phase control in the future.

In the present experiment, the nanotip samples such a small volume of the laser focus in comparison 
to the Rayleigh range and the focal waist size that it can be assumed that only a single intensity 
and phase are effective.
However, this is not the case in most ultrafast, laser-induced processes.
To account for the fact that common targets often experience a large range of laser intensities, 
one must properly weight the results with the intensity and target-density profiles, i.e.
the intensity-volume effect needs to be taken into account \cite{Sayler2007}.
Further, here we see that this problem is significantly exacerbated for broadband lasers as there
is an analogous and coupled phase-volume effect due to the two-dimensional phase profile.
In some situations, these complications could be reduced by physically limiting the extent of the
target or effectively limiting it by selecting a high-order process.
However, they are by no means eliminated.
For example, even for a target, very thin in the laser-propagation direction, $\Delta z \ll 
z_\mathrm{R}$, the phase is often strongly radially dependent even within the beam width.
Thus, the common assumption of an arctangent dependence of the CEP is often unwarranted and 
insufficient. 

Although these effects can complicate interpretation, they also have the potential to be used to
enhance desired effects and perhaps even be utilized in novel ways, for instance, to improve phase
matching in high-order harmonic (HHG) and attosecond pulse generation.
Further, one could optimize these effects by tailoring the interaction region via the input beam
geometry, expressed by $g_0$.
In addition, we expect that recent developments such as particle trapping and acceleration of atoms 
with femtosecond laser pulses may benefit from the CEP control demonstrated here through an 
interesting well-like structure formed by the CEP-gradients \cite{Shane2010, Freegarde1995, 
Eichmann2009}.
Hence, new ways of controlling atoms with large forces through ultrashort, strong laser fields may
result.
The detailed knowledge of the phase evolution found here impacts many fields where knowledge and 
control of the optical phase is vital and fascinating developments can be foreseen.

%%%%%%%%%%%%%%%%%%%%%%%%%%%%%%%%%%%%%%%%%%%%%%%%%%%%%%%%%%%%%%%%%%%%%%%%%%%%%%%%%%%%%%%%%%

% \bibliographystyle{naturemag}
% \bibliography{scibib}

% \section*{Supplementary information}
% is linked to the online version of the paper at http://www.nature.com/nature.

\section*{Acknowledgements}
We gratefully acknowledge Michael F\"{o}rster for supplying nanotips, and Peter Dombi and Sebastian 
Thomas for support in the measurement campaign.
This work has been supported by
the DFG Grant PA 730/5,
Laserlab-Europe EU-H2020 654148,
ERC Grant NearFieldAtto, DFG Cluster of Excellence Munich Center for Advanced Photonics.
D.H. acknowledges the Helmholtz Association for financial support.
M.K. acknowledges the Minerva Foundation and the Koshland Foundation for financial support.

\section*{Author Contributions}
All authors contributed to all parts of the experiment including the final version of the 
manuscript.

\section*{Methods}

We employ a phase-tagging method \cite{Johnson2011} for recording the CEP-dependent electron
back-scattering at tungsten and gold nanotips, placed at different points near the focus (see
Figs. S1 and S2 in the Supplementary information).
This method utilizes simultaneous measurements of electron time-of-flight spectra from the metal
nanotip in event-mode on the one hand, and of the CEP of every single shot  of the few-cycle pulse 
train on the other, the latter being determined by a phasemeter \cite{Wittmann2009}.
The events in both measurements are synchronized by triggering on the same laser pulse.
Both measurements rely upon the strong CEP dependence of backscattered, laser driven
photo-electrons, i.e. those electrons that return to their parent matter and scatter elastically 
back from it after acceleration in the optical field.

In the CEP-meter the two electron spectra that build up from xenon in the left and right
direction of the horizontal laser polarization are recorded and compared by calculating the
asymmetry parameter $A = \frac{N_L-N_R}{N_L+N_R}$, i.e. the contrast, where $N$ is the electron
yield left and right, respectively.
From this parameter the CEP of the laser shot can be evaluated \cite{Wittmann2009}.
Exploiting the high spatial resolution and pronounced CEP sensitivity of the electron
backscattering at the metal nanotip, we recorded phase-tagged kinetic-energy spectra at positions
on the optical axis and off-axis on hyperbolic curves in a plane along the focus (see Fig. 
\ref{fig:2DFScheme2} and S2 in the Supplementary information).

% \subsection{Data Availability}
% The data that support the plots within this paper and other findings of this study are available 
% from the corresponding author upon reasonable request.

\end{document}